\documentstyle[12pt,aasms4]{article}

\begin{document}

\title {Magnetically Driven Outflows in a Starburst Environment}

\author{Elisabete M. de Gouveia Dal Pino\altaffilmark{1,2}
\& Gustavo A. Medina Tanco\altaffilmark{1,3} }
\altaffiltext{1}{Instituto Astron\^omico e Geof\'{\i}sico, University 
of S\~ao Paulo, Av. Miguel St\'efano, 4200, S\~ao Paulo
04301-904, SP, Brasil; 
E-mail: dalpino@iagusp.usp.br, gustavo@iagusp.usp.br }
\altaffiltext{2}{
Astronomy Department,
Theoretical Astrophysics Center,
University of Berkeley,
601 Campbell Hall,
Berkeley, CA 94720-3411}
\altaffiltext{3}{Department of Physics and Astronomy, 
University of Leeds, EC Stoner Building, 
Leeds LS2 9JT, United Kingdom }

\begin{abstract}
 
We here investigate the 
 possibility that the observed collimated outflows 
in luminous infrared galaxies (LIGs) and some Seyfert galaxies
 can be produced
in a starburst (SB)
environment. In the former source class, 
in particular,
there seems to be some observational evidence for the presence
of nuclear SBs in some objects (e.g. Scoville, Lonsdale, \& Lonsdale 1998).

A nuclear disk can be quickly produced by gas infall during 
star formation in a $rotating$, stellar cluster. 
We find that massive nuclear SBs 
with core
disk masses 
$M_d \, \sim \,10^8 \, - 10^{9} \, M_{\odot}$,
and supernova rates 
$\nu_{SN} \simeq 5 \times 10^{-3} \, - \, 2 $ yr$^{-1}$
(which are consistent with the $\nu_{SN}$ values 
inferred from the observed non-thermal radio power in source candidates)
may inject kinetic
energies
which
are high enough to blow out directed flows from the
accreting disk surface, 
within the SB lifetimes.
In our models, the acceleration and 
collimation of the nuclear outflow are provided by magnetic fields 
anchored
into the rotating SB-disk. 
The emerging outflow carries a kinetic 
power that is only a small fraction (a few percent) of the 
supernovae energy rate produced in the SB.
Based on conditions determined from observed 
outflows and disks,
we find that moderate disk magnetic fields
 ($\gtrsim 8 \times 10^{-4}$ G)
are able to accelerate the outflows
up to the observed terminal velocities ($\lesssim$ few 100 km s$^{-1}$ in 
the case of the Seyfert galaxies, and
$ \sim$ 400 - 950  km s$^{-1}$ in the case of the LIGs).
The outflow is produced within a wind zone in the disk of radius
$\lesssim$ 100 pc  in the  the LIGs, and 
$\lesssim$ 10 pc in the Seyferts, with wind mass loss to disk accretion
rate ratios 
$ \dot M_w /\dot M_d \, \gtrsim \, 0.1$ (where
$\dot M_d \, \sim \, 100 \, M_{\odot} $ yr$^{-1}$). 
The observation of rotating nuclear disks of gas within 
$\sim$ few 100 pc scales,
like that in the prototype LIG, Arp 220 
(for which $\dot M_d \, \sim 100 \, M_{\odot} $ yr$^{-1}$),
and magnetized 
outflows in Sey galaxies and LIGs 
(with terminal $B \, \sim 10^{-5}$ G at the kpc-scales), 
provide some observational
 support for the magneto-centrifugal disk picture
drawn here. 
\end{abstract}

\section{Introduction}

Active Galactic Nuclei (AGN) emit large amounts of energy 
($10^{43- 48}$ 
ergs s$^{-1}$, from Seyfert galaxies  to QSOs)
over a large range of the spectrum coming from a small volume.    
The nature of the central engine that produces these extraordinary 
amounts 
of 
energy 
remains virtually unknown. Current AGN theories include: i) 
black holes (BHs) accreting matter from the surroundings (e.g. Rees 
1984);
ii) Starburst (SB) models in which activity is due to star formation 
in a central cluster (e.g. Terlevich et al. 1992 - who have 
revived old theories, e.g., of
Shklovskii 1960, Field 1964); and iii)
BHs surrounded by dense starbursts (e.g. Perry \& Dyson 1985, 
Scoville \& Norman 1988, Williams \& Perry 1994). 

While BH models (with masses $\sim \, 10^6 - 10^{10}
M_{\odot}
$ ) accreting matter from the surroundings
have the power to produce the observed luminosities and 
naturally provide  a preferred axis for collimating jets,
they seem to  explain unsatisfactorily, e.g., the 
broad emission-line regions (BELRs)  
of AGN
(as they require a fine-tuning to fit the observations). 
In Starburst models, on the other hand, 
strongly radiative compact supernova remnants (cSNRs)
are the source of the broad emission lines.  This model has been 
proved 
plausible in explaining the observed emission line spectrum,  the 
UV-optical 
continuum and related variability (e.g., Terlevich et al. 1992), 
but is 
difficult to be applied to the whole population of AGN. In 
particular, 
they do not provide a 
natural mechanism for jet production and collimation. We here 
address 
this question. 

It seems plausible that active star formation and a compact AGN  
activity may coexist in the nucleus of  
luminous infrared galaxies (the so called LIGs which
are characterized by extreme mid- to far-IR 
luminosities $L > 10^{11} L_{\odot}$ and high brightness temperature 
radio 
emission). The question to assess the relative contribution of each 
form of activity to the LIGs energetics, and to determine a 
potential 
evolutionary progression from one to the other has been recently 
addressed  by Smith, Lonsdale \& Lonsdale (1998, hereafter SLL98). 
Searching for a SB-compact object AGN connection in  these 
luminous infrared 
galaxies  they
have demonstrated 
that nuclear starbursts can in principle explain 
the observed VLBI- and VLA- scale structures and the  high infrared
 luminosities of most of these objects,
although in some cases severe constraints must be imposed on the 
initial 
mass function and, in general, complexes of extremely luminous radio
supernovae (RSNs) (clustered in pc-scale clumps) 
are required to explain the high-$T_b$ observed.
In a sample of 10 LIGs, they found potentially 6 SB systems, 3 
possible 
hybrid/intermediate systems and 1 quasar-like object (or BH candidate). 
This  would 
favor an evolutionary interpretation in which a
SB eventually begins to fuel a coalescing central compact object.
Galactic mergers are believed to be responsible for the origin of 
LIGs, and
numerical simulations (Mihos \& Hernquist 1996) suggest that
the SB occurs relatively late in the merging process ($\sim$ $10^8$ 
yr).
SLL98 results, in turn, suggest that a SB phase might carry on for 
another 
$\sim 10^8 $ yr as the nuclei coalesce and a compact AGN develops. 
Once 
the compact AGN fully turns on, it must quickly consume its shroud 
of dust
thus extinguishing the star formation process.

A prototype of the LIG class,  Arp 220, seems to be 
dominated by a massive
nuclear starburst imbedded in a rotating disk-like structure of 
dense molecular gas with a radius  $\sim 235 $ pc,
 and with densities ($n \sim 10^4$ cm$^{-3}$) and temperatures  
similar to those of the GMC
 core regions in our 
Galaxy (Scoville, Yun, \& Bryant 1997, hereafter SYB97). 
Several other systems (e.g., NGC 6240 and Mrk231) bear striking 
similarities 
to Arp 220, which are suggestive of a centrifugally supported disk of dense
molecular gas. These disks may play a critical role in the 
evolution of ultra-luminous galactic nuclei and in their energy 
release.
The high density of the gas will undoubtedly promote extremely high 
rates of massive star formation
since the conditions are very similar to those of GMC cores.
In these cases,  the presence of strong, highly 
ordered magnetic fields could provide both the conditions 
appropriate for RSN formation and production of 
collimated  outflows.

Nuclear outflows have been observed in  several Seyfert galaxies 
and LIGs
(e.g., Arp 220, NGC 6240, NGC 2782,  
NGC 4945, and NGC 4235) (see, e.g., Heckman, Lehrnet,
\& Armus 1993, Colbert et al. 1996,  Jogee et al. 1998, and references therein). 
In the case of the LIGs, the detection of emission line profiles 
with typical line widths of 200-600 km s$^{-1}$, and in some cases with 
FWHM as large as 1000 km s$^{-1}$ ( e.g. NGC 6240), are signatures of the 
presence of high-speed,
supersonic outflows where the lines are produced by shock-heated ambient gas 
which is swept by the outflow (Heckman, Armus, $\&$  Miley 1990, herareafter 
HAM90). In the case of the Seyfert galaxies, 
radio observations of a distance-limited sample of 
edge-on 
galaxies  by Colbert et al. (1996), have shown that 6 of 
the 10 
observed galaxies, have extended radio structures with radial 
extent  between 5 and 30 kpc. They are oriented at skewed angles 
($\Delta \, = \, 35^o  - 90^o $) with
respect to the galaxy minor axis  and 
do not look like the wide-angled winds observed in typical starburst 
galaxies.
Their morphologies resemble instead, those of more  collimated 
outflows 
originating in the nuclear region (within 1 kpc), like the jets  
observed in extended radio galaxies, although the Seyfert outflows  
are weaker 
(with  radio power at 4.9 GHz, $P_{4.9}\, = \, 10^{20-22}$ W Hz$^{-
1}$), 
smaller, and less collimated than those of the radio galaxies 
(which have typical radio power at 
1.4 GHz,  $P_{1.4} \, > \, 10^{24}$ W Hz$^{-1}$). 
Terminal outflow velocities
 of $\sim $ 25 to 
150 km s$^{-1}$ and magnetic fields $\sim 10^{-7}$ to $10^{-6}$ G at 
5 to 30 kpc
scales (implying average fields $\sim 10^{-5} $ G at $\sim $ 1 kpc)
 have been inferred from the observations.

Based on the evidences above, it seems plausible that the LIGs and 
perhaps  some of the low-level-activity AGN class of Seyfert galaxies 
may be powered by nuclear SBs at least during part of their active lives. In the 
present 
work, we assume that a SB, which has been induced
in an interaction or merging process, is in progress in 
the nuclear
region of these objects and  examine the conditions under which
collimated outflows can be produced in such a scenario. 
Previous purely hydrodynamic numerical studies of 
starbursts generated within few kpc radius from the nucleus 
have shown that SB mass loss dynamics may provide 
favorable conditions for a free vertically 
directed
escape of hot gas from the galactic environment 
at the vigorous SN dominated phase of the SB 
(e.g., Tomisaka \& Ikeuchi 1988, De Young \& Gallagher 1990, 
Suchkov et al. 
1994, Tenorio-Tagle \& Mu\~noz-Tu\~n\'on 1997). 
Those models have successfully explained the formation of the poor-collimated, 
 wide-angled super-winds (with radius $\sim$ 1 kpc)
 often observed in SB galaxies (e.g., HAM90).
In this work, however, we are mainly interested in systems with potential 
SB-AGN activity connection and thus examine the production of outflows 
within radial
distances
 of few tens of parsecs from the central core. We consider 
a mechanism that can provide efficient collimation and acceleration 
of the emerging outflow on these scales. 
We assume that
part of the gas released by star formation in the rotating SB system 
is quickly collected 
into a disk with similar conditions, for example, to those observed 
in the Arp 220 nuclear disk. Observations 
support the evidence for magnetized outflows. Since magnetic winds 
can
be produced by rotating objects with a magnetic field anchored into 
them
(e.g., Weber and Davis 1967), we assume that magnetic fields are 
present due to magnetic-flux captured from the environment during 
the formation of the disk, and  compression by the 
accretion process (see, e.g., Spruit 1996).

The perception that rapid rotation coupled with
strong magnetic fields could drive outflows was first
realized in the context of the solar wind theory (e.g., Weber 
and Davis 1967, Mestel 1968).
Later, Blandford $\&$ Payne (1982) introduced a scale 
free
hydromagnetic wind model for jet acceleration from a magnetized
disk which gives centrifugally driven outflow solutions.  These 
solutions also have the possibility of being self-collimated 
(Heyvaerts $\&$ Norman 1989).
(See, however, Okamoto 1998 for a discussion on the limitations 
of these models.) 
Similar outflow models were subsequently applied to proto-stellar 
objects 
(e.g., Pudritz \& Norman 1983, Pelletier \& Pudritz 1992, 
 Wardle \& K\"onigl 
1993, Shu et al. 1994, see also Spruit  1996, for a review), 
galactic X-ray
binaries (e.g., Blandford 1993), and
proto-jovian planets (Quillen \& Trillen 1998).
In the present work, we use the Pelletier and Pudritz solutions 
(hereafter
PP92) to investigate the formation of magneto-centrifugally 
accelerated outflows in a SB environment.

In \S 2 we try to build a consistent SB scenario and examine
whether massive nuclear star formation may be present at high enough 
rates to drive  galactic flows out from the 
surface of the 
nuclear gas disk produced in the course of the SB evolution. 
In \S 3, 
we discuss the physical conditions for the nuclear disk to 
produce collimated outflows based on the observed characteristics of 
the
Arp 220 nuclear disk, and estimate the 
radius,
accretion and wind outflow rates, terminal velocities, kinetic energy rates, and 
magnetic field ranges 
for which 
magneto-centrifugal outfows 
could operate. In \S 4, we draw our conclusions and 
briefly discuss our results.

\section{A Starburst Scenario}

In an attempt  to 
build consistent starburst 
scenarios and evaluate their energetics, mass loss, and 
rate of supernova production, 
in previous work (de Gouveia Dal Pino \& Medina Tanco 1997, hereafter GT97) 
we have calculated the evolution of 
several starburst stellar clusters based on the evolutionary 
population synthesis models of Schaller et al. (1992). 
As initial conditions we have 
assumed a Salpeter initial
stellar mass distribution
function [$\Phi(m) \,  \propto \, m^{-\gamma}$ ],
with a spectral index $\gamma \, = 2.35$, and 
stellar masses in the range [M$_l$ = 1 - 8
M$_{\odot}$,
M$_u$ = 120 M$_{\odot}$]; 
initial number of stars $N_c \, = 10^6 - 4 \times 10^9 $ 
(corresponding to cluster radii $R_c \, \simeq 0.2 - 705 $ pc
and masses M$_c$  $\simeq$ 10$^7$ - 10$^{10}$  M$_{\odot}$); 
and an energy deposited by each supernova (SN) 
$E \, = 10^{51}$ erg s$^{-1}$.
Assuming also that the system was rotating with an ellipticity 
$\epsilon \, = \, J/r_c v_s \, \simeq \, 2.5 \times 10^{-3}$,
where $J$ is the angular momentum per unit mass (Salpeter and Saslaw 
1996, Begelman \& 
Rees 1978), we have found that part of the gas released by star 
formation may cool quickly by free-free emission in 
time scales much smaller than the evolutionary time scales of the 
system ($\tau \lesssim \, 10^8$ yr) 
and thus fall towards the centre and
settle into a disk with an initial 
characteristic radius
$r_d \, \simeq \, \epsilon ^{1/2} r_c$.
In fact, rapid fueling is expected by the large star 
formation rates and gas
conversion 
into stars detected in nuclear SB regions 
(Kennicutt et al. 1987, Lo et al.
1987). These observations indicate that 
fueling should occur in time scales smaller than the conversion 
time into stars ($\sim \, 10^{7} - 10^{8} $ yr)
if the gas is to accumulate faster than it is
consumed (e.g., Tenorio-Tagle \& Mu\~noz-Tu\~n\'on 1997).
As an example, 
Figure 1 (see also GT97) shows the results for a cluster with
3 $\times $ 10$^8$ M$_{\odot}$
 and a core disk which eventually becomes also gravitationally 
unstable and undergoes
star formation (see below). For a cluster-disk system with masses 
and radii about ten times 
larger (smaller) than those of Fig. 1,
all the luminosity values and supernova rates increase (decrease) by 
an order of magnitude.

As investigated in previous SB-driven outflow models, a strong nuclear SB can, 
in principle, 
drive a wide-angled wind
out of the system  
(see e.g., HAM90).
We will first discuss this process  in terms of the
energy available to produce 
winds in a SB system and then,
on subsequent sections, examine the mechanism for providing outflow 
acceleration and collimation.
If an outflow is driven by a SB, then the observed 
non-thermal radio power ($P_{nt}$) should be correlated with 
the  supernovae production rate ($\nu_{SN}$)  
in the SB  through
 (Condon \& Yin 1990):
\begin{equation}
P_{nt} \, \simeq \,  1.3 \times 10^{23} \, {\rm W Hz}^{-1} \,
\nu_{SN} \,
\left({ \nu \over 1 \, {\rm GHz} }\right)^{-\alpha_r} 
\end{equation}
\noindent
where $\nu$ is the observed radio frequency, and $\alpha_r 
\, \simeq \, 0.7-0.8 $ is the 
spectral index.  Using the observed radio power for the Seyfert 
outflows 
we find  (see  Condon et al. 1996)
 $\nu_{SN} \, \simeq \, 5.1 \times (10^{-3} - 10^{-1}$) yr $^{-1}$,
which is consistent with the predicted values of
$\nu_{SN}$  by the SB models (see, e.g., Fig. 1).

Alternatively, simple analytical prescription can be used to 
estimate the
expected  supernova production rate $\nu_{SN}$ as a function of the 
star formation rate (SFR). For a power law initial 
stellar mass function, with upper and lower stellar mass
cutoffs 
  $m_u \simeq 120 M_{\odot}$, and $m_l \simeq 1 
M_{\odot}$, respectively,
we have (e.g., Scoville \& Soifer 1991) 
\begin{equation}
\nu_{SN}  = \int { \Phi(m) dm } \,
\simeq { SFR \over 3 } \,
{\left({m_{SN}^{-1.35} - m_u^{-1.35} }\right) \over
\left({m_{l}^{-0.35} - m_u^{-0.35} }\right) }  
\end{equation}
\noindent
where $m_{SN} = 8 M_{\odot}$ is the minimum mass for supernova
detonation, or
 \begin{equation}
\nu_{SN}  \simeq 0.5 \, yr^{-1}
 \left(SFR \over {20 \, M_{\odot} \, yr^{-1} }\right) 
\end{equation}
\noindent
which shows that SFR $\simeq (5 - 20) \, M_{\odot} \, yr^{-1}$ 
result SN rates consistent with those required to
explain the 
observed radio power in Sey outflows. 

In the case of the LIGs, the observed infrared luminosities
imply star formation rates  
SFR $\simeq$ 16 - 125 $M_{\odot}$ yr$^{-1}$.
SBs models with these SFR 
result in supernova rates 
 $\nu_{SN} \simeq $ 0.1 - 2  yr$^{-1}$ (SLL98).
These are also consistent with the 
$\nu_{SN}$ values required
by the observed non-thermal radio luminosity in these
objects. In the particular case of the Arp 220 source, 
the supernova
rate required by the observed non-thermal radio power is
$\nu_{SN} \simeq $  2 yr$^{-1}$ (which is in agreement with
the predicted rate by a SB model with cluster and disk 
masses an order of magnitude
larger than those shown in Fig. 1, 
$M_d \, \simeq \, 10^9 M_{\odot}$).

SBs with 
$\nu_{SN}$ $\simeq$  5 $\times$ 10$^{-3}$  - 2 yr$^{-1}$
may inject $kinetic$ energy into  the ambient medium 
at a rate  (see, e.g.,  Eq. 1; HAM90)
\begin{equation}
L_K \, \simeq 3.5 \times 10^{43} \, \nu_{SN} \, \simeq \, 
(2 \times 10^{41} -  7 \times 10^{43}) \, 
{\rm erg  s}^{-1} 
\end{equation}
\noindent
which is  comparable to both the rate of 
gravitational energy released by the infalling gas 
 and the SN luminosity in the core disk of the SB models (see, e.g.,  Fig. 
1, for which $\nu_{SN}$ $\lesssim$ 0.2  yr$^{-1}$, and 
$L_{SN} \, \simeq \,  L_{K} \, \simeq \, 10^{43}$ erg s$^{-1}$),
and smaller than  ($\lesssim$ 20 \%)   the observed 
bolometric luminosities of the sources
($L_{bol} \simeq L_{FIR} \gtrsim 10^{11} L_{\odot}$ 
for typical LIGs and Sey 
galaxies).
In fact, we will see later (\S 3) that only a fraction of this 
SN energy rate is actually necessary to power the outflow. The remaining SN 
energy will be possibly lost in radiation, as indicaterd by recent numerical 
studies (Thornton, Gaudletz, Janka, \& Steinmetz 1998)
 
\section {Modelling the Outflow}
\subsection { Disk Parameters}
In the previous section, we have assumed that a rotating nuclear
disk of gas is produced during the SB evolution and have evaluated 
the amount of $kinetic$ energy that such a system is able to inject 
into the ambient medium.
Now, before deriving the conditions for outflow 
acceleration and collimation from the surface of the disk, we 
 must explore fiducial physical parameters for the disk.
 In earlier studies of  SB-driven, wide-angled super-winds (e.g., HAM90), 
global physical conditions were derived for the 
central regions (radius $\lesssim$ 1 kpc) 
of SB galaxies. Here, since we are focusing
on the inner  regions (located within  
few tens of pcs radii from the 
nucleus) of systems with potential coexistence of a SB-AGN activity, we take the 
observed data of  
 the nuclear disk of the  LIG prototype, Arp 220 
(SYB97).  
High-resolution CO imaging  and dust continuum emission
of the nuclear region of Arp 220 has revealed the
presence of a rotating disk feature of molecular gas with mean
density 
$n_{H_2} \simeq$ 2 $\times$ 10$^4 $ cm$^{-3}$, 
a radius
$\simeq$ 250 pc, and   two 
stellar nuclei at $\sim $ 235 pc from the center. 
Most of the disk
mass resides in the
molecular gas, rather than in the stars, and
most of the molecular gas 
(M$_d $ $\sim$  5.4 $\times$ 10$^9$ M$_{\odot}$)
 is
concentrated in a very thin disk within 
the double nuclei distance 
from the center.
The fact that the bulk of the molecular gas has relaxed into a disk with large 
mass of gas concentrated interior to the double stellar nuclei is consistent 
with a scenario in which the gas in merging systems settles into the center 
faster than the stellar components of the SB.
The mass in the Arp 220 disk
is approximately distributed according 
to the empirical relation
(see Fig. 8 of SYB97)
\begin{equation}
M_r \simeq 5 \times 10^9 \, M_{\odot} 
\left({ r \over 200 \, {\rm pc} }\right)^{\beta}
\end{equation}
\noindent
with  $\beta \simeq$ 1.66 (for $r \lesssim $ 200 pc).

The rotation velocity is related to the mass distribution through
the simple spherical approximation 
$v_{rot} = (G M_r/ r)^{1/2}$.

For a self-gravitating disk, the effective thickness (H) of the 
gas can be obtained from the standard condition of hydrostatic
equilibrium (Spitzer 1942),
$H \, = \, 
\left({ 0.233 \, \sigma^2 \over G \, \Sigma}\right) $,
where $\sigma \simeq$ 90 km s$^{-1}$ is the vertical dispersion
velocity (SYB97), and $\Sigma = M_d/A_d = 2H \rho_d$
 is the mass surface
density of the disk. Within a radius r $\simeq$ 200 pc and 
$M_d$ $\simeq$ 5 $\times$ 10$^9$ M$_{\odot}$, 
$\Sigma \simeq 3 \times 10^4$ M$_{\odot}$ pc$^{-2}$ 
for Arp 220, so that
\begin{equation}
H \simeq 15\, {\rm pc}
\left({ \sigma \over 90 \, {\rm km \, s^{-1} }  }\right)^2
\left({ \Sigma \over 3 \times 10^{4} \, M_{\odot} \, {\rm pc^{-2} } 
}\right)^{-1}.
\end{equation}
This implies a thin disk at r $\simeq$ 200 pc with 
$H/r \, \simeq \, 0.075$.  In the inner region it can be thicker
however.

The stability of the gas disk to axi-symmetric perturbations has been 
examined by SYB97 who have used the Toomre (1964) criterion
$Q \simeq \sigma k/ (\pi G \Sigma)$ (where $k \simeq 2 \Omega $ 
is the epicyclic frequency). The fact that Q $\lesssim $ 1 
within r $\simeq$ 400 pc implies that the gas disk is 
unstable and will develop large density enhancements, 
ultimately leading to the formation of massive star clusters in 
the disk. This result is consistent with our previous 
assumption of a disk originated in a SB environment also 
processing star formation.

The extreme gas densities in the central disk of Arp 220
($n_d = n_{H_2} \simeq 2 \times 10^4 $ cm$^{-3}$)
 will 
 result in high accretion rates. Assuming 
a viscous transport of angular momentum,
SYB97 have used the Pringle (1981) treatment to 
estimate the rate of gas accretion as a function of the
disk radius. For the typical Arp 220 parameters, they found
accretion rates 
$\dot M_d \, \simeq $ 100 $M_{\odot}$ yr$^{-1}$ peaking
at $\sim $ 250 pc
(which corresponds to a kinematic viscosity 
$\nu \simeq 3 \times 10^{27} $ cm$^2$ s$^{-1}$).

We now must check whether the above accretion rate $\dot M_d$ 
is consistent with
the range of  temperatures and pressures  estimated for
the disk.  Accretion through the disk 
implies that there is dissipation of energy into the disk.  
For a self-luminous disk the temperature as a function of
the  radius is
${T_r}^4 = {3 G M_d \dot M_d \over 8 \pi \sigma_{SB} r^3}$,
where $\sigma_{SB}$ is the Stefan-Boltzmann constant.  Using the 
accretion
rate mentioned above we estimate
\begin{equation}
T_d \simeq 14 \, {\rm K} \,
\left({ M_d \over 5 \times 10^9 \, M_{\odot}}\right)^{1/4}
\left({\dot M_d\over 100 \, M_{\odot} \, {\rm yr^{-1} }}\right)^{1/4}
\left({ r\over 200 \, {\rm pc }}\right)^{-3/4}
\end{equation}
\noindent
which is consistent with the  temperatures estimated
from the OH observations in Arp 220 ($\simeq $ 17-21 K; SYB97).

The sound speed at this radius would be
$v_s \sim 0.4 \, (T/14 \, K)^{1/2} $ km s$^{-1}$, and the disk 
pressure
\begin{equation}
P_d \simeq 4 \times 10^{-11} \, {\rm dy \, cm^{-2} } 
\left({ n_d \over 2 \times 10^4  \, {\rm cm^{-3} } }\right)
\left({ T \over 14 \, {\rm K} }\right).
\end{equation}
\noindent
We note, however, that in our SB/disk model, massive star formation may still 
 be occurring after the formation of the disk and even in the disk itself. In 
such a case, one would expect the presence of an ionized, warmer gas component 
produced in interactions between the ejected matter in the SB (e.g., HAM90). A
 disk with a dominant warm gas component (with T$_d$ $\simeq $ 
10$^{4}$ K) 
would have a higher pressure 
$P_d \, \simeq  \, 10^{-8}$ dy cm$^{-2}$.
In any case, no matter whether the inner disk is cold (i.e., presently not 
undergoing massive star formation) or warm, the magneto-centrifugal mechanism 
for outflow production that  will be discussed below is applicable to both.

\subsection{ Outflow Conditions}

Previous SB-driven modeling of wide-angled super-winds
(e.g., Tomisaka \& Ikeuchi 1988, De Young \& Gallagher 1990, 
Suchkov et al. 
1994)
have shown that the multiple action of stellar winds and SN explosions
may lead  to the formation of a shell of shocked disk material which, 
after blowout, is dragged to the halo, where it forms the walls of a 
bi-conical cavity. This would, ultimately, create conditions for a free 
directed
escape of hot gas at the later, more vigorous SN dominated phase of the SB.
Formulation derived by Mac Low \& McCray (1988) and Mac Low, McCray, \& 
Norman (1989) gives a simple condition for a SB-driven outflow to blow out of the 
atmosphere above the disk. The rate of energy released by the SN, L$_K$ (eq. 4), 
must 
 satisfy 
\begin{equation}
L_{K} \, > \, 3 \times 10^{5} \, L_{\odot}
{\left({ n_d \over 2 \times 10^4 cm^{-3} }\right) }^{-1/2} 
{\left({ H \over 15 \, pc }\right) }^{2}
{\left({ P_d \over 10^{-8} \,  dy \, cm^{-2} }\right) }^{3/2}
\end{equation}
\noindent
which is a condition more than satisfied by the 
candidate systems, which have
 $L_{K} \simeq (5.3 \times 10^{7} \, -  \, 2 \times 10^{10}) \, L_{\odot}$ (eq. 
4).
We note that this condition still holds 
 for more compact, denser disks with, for example, 
$H \sim $ 1 pc, n$_d \, \lesssim \,  10^8 $ cm$^{-3}$, and P$_d $
$< $ 10$^{-3}$ dy cm$^{-2}$.


The results above suggest that nuclear SBs may 
satisfy  the conditions for driving winds out of the galactic core 
of the 
Seyferts and LIGs,
but, how to explain the collimated morphology  of  the observed 
outflows  that we are investigating and their acceleration?
One could invoke a thick gas torus surrounding the core  to 
collimate the 
"wind" in a similar way to that proposed in standard models for AGN 
(see, e.g.,
Antonucci 1993). As mentioned earlier, 
hydrodynamic models for  SB super-winds 
predict the formation of a cavity that provides a natural channel for 
vertical escape of the gas ejected by the SN.
Recently, Tenorio-Tagle \& Mu\~noz-Tu\~n\'on (1997), have 
explored hydrodynamic solutions in which the rapid fall of 
disk matter towards 
the
center develops a shock front in the inner region of the disk, 
and the material 
crossing it experiences a splash and is ejected vertically to 
the lower pressure 
zones above the disk, thus forming a (poorly collimated)
kpc-long cone. 
All these models seem to be able to provide appropriate conditions for the 
creation of the observed, poor-collimated super-winds emerging from the kpc-scale 
galactic disks of SB galaxies.
Here, in our investigation of the dynamics of nuclear 
SB/disks, we examine another  
 potential  mechanism 
in which efficient collimation and outflow acceleration
is provided by 
magnetic fields anchored into the core disk of the SB
$-$ an 
assumption that is supported by the observation of magnetic fields
in the outflows of the Seyferts 
(B $ \simeq 3.5 \times 10^{-7} - 1.6 \times 10^{-6}$ G at 
few kpc above the disk; Colbert et al. 1996). 
Such mechanism could also, in principle, be operating in the inner (possibly 
unobservable) regions of wide-angled super-winds of normal SB galaxies.

In a  magnetized gas
under low ionization fraction condition, 
for an outflow to be launched the neutral component must
be well coupled with the ions (e.g. Wardle \& K\"onigl 1993).
This coupling is
described by $\eta$, the ratio of the dynamical time to the mean
collision time of a neutral atom in a diluted sea of ions 
(e.g. Wardle \& K\"onigl, Shu et al. 1994, 1995),
$
\eta = {\gamma_i \rho_i r \over v_{rot}},
$
where 
$\rho_i$ is the density of ions and 
$v_{rot}$ is the rotational velocity of a particle in a circular 
orbit.
When $\eta \gg 1$ the neutrals move with the ions and
so can be described as well coupled to the magnetic field.
Here 
$\gamma_i \sim 3 \times 10^{13}$ cm$^3$ g$^{-1}$ s$^{-1}$
is the drag coefficient between ions and neutrals
and $\rho_i$, the density of ions,
is the product of the atomic mass of the ions,
the ionization fraction and the total density.
We can then estimate
\begin{equation}
\eta > 5.6 \times 10^2 
\left({n_e/ n_H \over 10^{-5}}\right)
\left({\rho_d \over 10^{-19} \, {\rm g~cm}^{-3}}\right)
\left({M_d \over 5 \times 10^{9} \, M_{\odot} }\right)^{-1/2}
       \left({r \over 200 \, {\rm pc}}\right)^{3/2}
\end{equation}
\noindent
For clouds with temperatures, abundances and column densities not 
much different from those in the disk examined here, Fria\c ca \& 
Jafelice (1998) have found that the ionization state is roughly inversely 
correlated with the density. From their curves we 
have estimated the corresponding value of $n_e/n_H \, \simeq \, 10^{-5}-10^{-2}$ 
for our density and temperature ranges. 
 We conclude that over the whole disk $\eta \gg 1$ and the
neutrals can be said to be well coupled to the ions.
However this does not mean that the particles are necessarily
well coupled to the magnetic field.

The magneto-centrifugal wind models can operate where the 
charged particles are sufficiently coupled to the field lines that 
they
are flung out centrifugally from the outer layers of the disk
(for a favorable field geometry; see below).   This occurs when
the magnetic Reynold's number is $Re_M \, \gg 1$.
We can estimate the  magnetic Reynolds number as 
$Re_M = {v_{rot} H \over \nu_M}$,
where $\nu_M$ is the magnetic resistivity,
$\nu_M = c^2/(4 \pi \sigma_e) $, 
and $\sigma_e $ is
the electric conductivity determined from 
collisions between electrons and H$_2$ molecules, 
$\sigma_e \, \simeq  \, {n_e e^2 \over n_{H_2} m_e < \sigma_c v > }
\, \simeq  \, 1.1 \times 10^{16} (n_e/n_H) $ s$^{-1}$
(where $\sigma_c$ is the cross section for $e-H_2$ collisions,
$v$ is the relative velocity between them, and $< \sigma_c v> $
is the Maxwellian mean of the product;
Hayashi 1981, Takata \& Stevenson 1996). 
Using the Arp disk parameters we estimate
\begin{equation}
Re_M > 2.3 \times 10^{18} 
\left({ M_d \over 5 \times 10^{9} M_{\odot} }\right)^{1/2}
\left({H \over 15 \, {\rm pc} }\right)
\left({n_e/n_H \over 10^{-5} }\right)
\left({r \over 200 \, {\rm pc}}\right)^{-1/2}
\end{equation}
\noindent
so that 
we can assume that the magnetic field is well coupled to the gas.

\subsection{Magneto-centrifugally Driven Outflows} 

Collimated winds may be produced by a rotating disk with a magnetic 
field (${\bf B}$) anchored into it (see, e.g., Spruit 1996 and 
refereces therein).
The magnetic field and disk structure will determine 
how much mass is launched out of the disk.

As remarked before, we here assume that the field is 
produced due to magnetic 
flux capture from the environment during the formation of the
disk in the SB, followed by advection and compression 
by the accretion process.
 Poloidal
flux capture in this way cannot be destroyed by local processes 
in the disk, it 
can only escape by diffusing radially outward (e.g., Spruit 1996). It is out of 
the scope of this work to formulate a theory for (turbulent) 
diffusion in the 
accretion disk, so that we cannot reliably predict the distribution
 of the 
magnetic field. However, since all the energy densities are expected 
to increase towards the center,
it is reasonable to assume that the field will increase 
inward as well 
and form an opening field geometry. As argued by  Spruit, Stehle, \&
Papaloizou  (1995), fields 
with such a configuration can be quite strong, 
with magnetic energy densities 
exceeding the gas pressure, which must make them ideal 
for magnetic outflow 
production.

Just above the disk, the gas density
is typically low enough, so that the magnetic energy density is 
large
compared with the thermal and rotational energies. The field 
in this region may be thus assumed to be force-free (${\bf J x B  
= 0} $). The poloidal magnetic field 
lines that raise from the disk are somewhat like rigid wires that
control the flow. 
The gas of  the disk surface will be forced to 
corrotate with the field lines  raising out of the disk. 
For a sufficiently cold disk, the solely forces acting on a 
parcel of gas will 
be the gravitational force and the centrifugal force due to rotation. 
Along the field line above the disk, the centrifugal 
force will 
increase 
with the distance from the axis and the effective gravity decrease. 
When the 
centrifugal force component along the line exceeds that of gravity, the gas 
tied to the field line is accelerated outward. This outward centrifugal 
acceleration continues up to the Alfv\'en point [the location where
the outflow poloidal speed reaches the Alfv\'en speed, $v \, = \, v_A \, = \, 
B^2/(4 \, \pi \,  \rho)^{1/2}$, and  stops when the flow attains the fast 
magnetosonic point. At that 
point,
the field is no longer strong enough to enforce corrotation.
Beyond the Alfv\'en surface ($\rho v^2 \, > \, B^2/8\pi$),
the inertia of the gas causes it to lag behind the rotation of 
the field line, and the field winds up
thus developing a strong azimuthal component ($B_{\phi}$). 
The 
magnetic stress force associated to  the azimuthal
field  [$-B_{\phi}^2/(4 \pi r$), 
where r is the radial distance from 
the disk axis]
will provide the flow collimation (see e.g., PP92, Spruit 1996).

A self-gravitating accretion disk with an accretion rate $\dot M_d$
releases a known amount of gravitational binding energy of gas as it
spirals inward to the center. The presence of viscous stress 
causes the gravitational binding energy to transform into heating
and the angular momentum is radially transported outward by 
turbulent
stress. In the presence of magnetic fields, 
however, part of the 
gravitational energy released in the accretion is converted into 
mechanical energy of a wind. In addition, the angular momentum 
is efficiently transported by the wind through the disk surface,
and the wind mass-loss rate $\dot M_w$ has only to be a tiny 
fraction of $\dot M_d$ (see, e.g., PP92 and references therein).
Using the disk angular momentum equation one can show that
as long as the magnetic pressure is comparable or larger than the
gas thermal pressure, the magnetic torque will completely
dominate the viscous torque (e.g. PP92). In fact, the ratio of 
the magnetic to the viscous torque is
\begin{equation}
 \left({ \tau_w \over \tau_{vis} }\right) =
       \left({ B_z^2 \over {4 \pi P_d} }\right)
       \left({r_A \over \alpha_{SS} H }\right)
\end{equation}
\noindent
where $\alpha_{SS} < $ 1  
is the Shakura-Sunyaev viscous prescription (e.g., 
Frank, King, \& Raine 1992),
and $r_A$ is the Alfv\'en radius. Thus, since 
$r_A/\alpha_{SS} H > 1$, and the
magnetic pressure in the disk 
($B^2/8 \pi \gtrsim 2.5 \times 10^{-8} $ dy cm$^{-2}$; see below)
is effectivelly of the order or 
 larger than the thermal pressure
 (Eq. 8), we expect that the angular momentum 
loss through the wind will dominate.

In  magneto-centrifugal disk outflows, 
the angular momentum of the wind is tied to that of the disk, 
and so the wind flux is directly related 
to the accretion rate, $\dot M_d$, through the disk.
To investigate 
specific conditions under which collimated winds could be generated 
in the SB/disk scenario drawn in the previous paragraphs,  
we here follow the non-self-similar,
steady-state, magneto-centrifugally 
driven wind models developed by PP92.
 In their solutions, the angular velocity of the disk
is assumed to be a power law of the magnetic flux, 
$\Omega \propto \psi^{-\alpha}$. 
They have found that for
a power law index  $\alpha = 3$,
the outflow carries a constant current intensity everywhere and
is free of the divergence found in self-similar models (e.g.,
Blandford \& Payne 1982). 
We here adopt this  relation. Now, according to
Eq. 5,  the angular 
velocity scales with the disk radius as
\begin{equation}
\Omega = { (G M_r )^{1/2}\over r^{3/2} }  \, \propto  \, r^{(\beta - 3)/2}
\propto \, r^{-0.67}.
\end{equation}
\noindent
Therefore, the magnetic flux will also be a power law 
of the radius,
\begin{equation}
\psi \propto r^{(3 -\beta) \over 2 \alpha}
\, \propto \, r^{0.223}.
\end{equation}
This, in turn, implies
the following relation for the poloidal 
magnetic field in the disk as a function of the radius (PP92), 
$B_r \, \propto r^{-\delta}$,
with $\delta = 2$, or
\begin{equation}
B_r = B_{\infty} 
\left({r_{\infty}  \over r }\right)^{\delta}
\end{equation}
\noindent
where $B_{\infty} \simeq 10^{-5} $ G is the average value
inferred from the observations of the Sey outflows at 
the kpc-scale, $r_{\infty} = 1000$ pc.

In these models, the ratio of the radius $r$ 
to the Alf\'ven radius, $r_A$,  ${1\over 3}(r/r_A)^3$, 
can be determined by the  parameter (PP92)
\begin{equation}
\epsilon \equiv {\dot M_d \sqrt{G \, M_r} \over 3 B_r^2 \, r^{5/2}}
\end{equation}
\noindent
For a wind to be produced the field lines must be
more than $20^\circ$ from normal to the disk and $\epsilon \leq 
1$.
Assuming $\dot M_d$ constant,
the relation above for B$_r$ (Eq. 15) and Eq. 5 for the disk 
mass result
\begin{equation}
\epsilon \propto r^{ {(\beta - 5) \over 2} + 2 \delta} 
\propto r^{2.33}. 
\end{equation}
Using these relations and the physical conditions
in the Arp 220 disk 
(with constant $\dot M_d \, = \, 100 \, M_{\odot} \, yr^{-1}$), 
we can  determine a 
maximum radius $r_e$ for which $\epsilon(r_e) = 1$,
\begin{equation}
r_e \simeq 127 \, {\rm pc} \,
{\left({\dot M_d \over 100 \, M_{\odot} \, {\rm yr^{-1}} }\right)^{-0.43} }
{\left({B_{\infty} \over 10^{-5} \, {\rm G} }\right)^{0.86} }
{\left({ r_{\infty} \over 1 \, {\rm kpc} }\right)^{1.72} }
\end{equation}
\noindent
This value determines a characteristic maximum
radius for the outflow zone in the  disk.  We note, however, that it is 
sensitive to the adopted value of the magnetic field on
the large scales $B_{\infty}$ which is only
barely known. 
(Notice that Eq. 15 implies
$B(r_e) \simeq 6 \times 10^{-4}$ G on the disk surface at $r_e$.)

 For these models
the mass flux in the wind, $\dot M_w$, is related
to the accretion rate through the disk by (PP92)
\begin{equation}
\dot M_w = f \dot M_d
\end{equation}
where 
\begin{equation}
f \approx \left({r \over r_A}\right)^2.
\end {equation}

HAM90 have estimated   values for $\dot M_w$ for  wide-angled super-winds
observed in far-infrared galaxies ranging from a few $M_{\odot}$ yr$^{-1}$ in the
low-power sources (e.g., M82), up to $\simeq$ 50 $M_{\odot}$ yr$^{-1}$  in
powerful sources (like Arp 220 and Mrk 273). These values correspond  to the 
total amount of mass loss rate integrated over the whole  radius  
 of those winds. In the inner, few tens of pc scales, $\dot M_w$ is expected to be
smaller. In fact, in our model, $\dot M_w$ approximately scales 
as
$\dot M_w$  $\propto \, \psi^{2}$ (eq. 6.7 of PP92), or according to Eq. 14 above,
$\dot M_w$  $\propto \, r^{045}$.
Therefore, the value
 $\dot M_w$  $\simeq$ 50 $M_{\odot}$ yr$^{-1}$ 
 for the Arp 220 outflow will correspond to 
  disk radius $> >$ 100 pc . We 
 can then assume that in the inner few pcs
 $\dot M_w \, \simeq \, $ 10 $M_{\odot}$ yr$^{-1}$ and take 
 $\dot M_w/\dot M_d \simeq 0.1$.
 This value 
is compatible with estimates for 
YSOs and AGN outflow models (see, e.g. PP92 and references therein).  
Substituting this value into Eqs. 19, and 20 above, 
we can then estimate a fiducial value for the ratio
$r_A/r_i \, \simeq \, (\dot M_w/\dot M_d)^{-1/2} \, \simeq 3.2$
at a fiducial radius $r_i$,
and according to Eq. 16,
\begin{equation}
\epsilon_i  \simeq {1 \over 3}
{\left({r_i \over r_A}\right)^3 }
\simeq 1.1 \times 10^{-2}
\end {equation}
\noindent
Thus substituting this value into Eq. 17,
\begin{equation}
r_i = r_e \, 
\left( {\epsilon_i \over \epsilon_e}\right)^{0.43} \,
\simeq \,  18 \, {\rm pc}, 
\end {equation}
or (according to Eq. 16)
\begin{equation}
\epsilon_i \sim 1.1 \times 10^{-2}
\left({\dot M_d \over 100 \, M_{\odot} \, {\rm yr}^{-1} }\right)
\left({M_{r_i} \over 9.2 \times 10^7 \, M_{\odot} }\right)^{1/2}
\left({B_{r_i} \over 3.1 \times 10^{-2} \, {\rm G} }\right)^{-2}
\left({r_i  \over 18 \, {\rm pc} }\right)^{-5/2}.
\end{equation}
\noindent


 The terminal speed of the outflow can be evaluated from the solution of
 the Bernoulli equation for the flow in the asymptotic domain
 (e.g., PP92, eq. 6.6) 
\begin{equation}
v_\infty (r) \simeq \,
{ (2  \lambda)^{1/2} \over 3 \, \epsilon(r_e) } \,
(\Omega_e r_e)  \,  \psi^{-2} \, \propto \, r^{-0.45}
 \end{equation}
where ($ \Omega_e r_e$) is the rotation velocity in the disk
at the outer outflow radius $r_e$, and 
$\lambda$ is a constant everywhere in the wind
that can be estimated from Eq. 6.7 of PP92, and Eq. 14 
above
\begin{equation}
\lambda \sim  {4.5 \over 2} \epsilon(r_e)^2 
{\dot M_d \over \dot M_w(r_e) - \dot M_w(r_i)}
\left[ {1 - 
\left({r_i \over r_e }\right)^{0.45} } \right]  \, 
\simeq 9.35 
\end{equation}
where we have used the approximate scaling law for 
$\dot M_w(r) \, \propto \, \psi^2 \, \propto \, r^{0.45}$.
Substituting  $\lambda$ into Eq. 24,
we find the  expected terminal velocity at the radius $r_i$ at which most of
the wind originates 
\begin{equation}
v_\infty(r_i) \simeq \, 973 \, {\rm km \, s^{-1} }
\left({\Omega_e \, r_e  \over 282 \, {\rm km  s}^{-1}  }\right)
\left({r_i/ r_e \over 0.14}\right)^{-0.45}
\end{equation}
and $v_{\infty}(r_e) \, \simeq$  0.42 $v_{\infty}(r_i) \simeq 408 $ 
km s$^{-1}$.
These estimates are consistent with 
the detection of emission
lines typically with  widths  $\sim 200-600 $ km s$^{-1} $ in LIGs.
They 
indicate the presence of 
a high speed pre-shocked wind gas that was possibly decelerated to these velocities.
In the case of Arp 220, the observation of line widths declining 
from $\sim 700 $ km s$^{-1} $  
 near the axis of the outflow to $\sim 300 $ km s$^{-1} $  at the outer 
 radius are also consistent 
 with the prediction of our model of a decreasing  outflow velocity with radius.
On the other hand,  the  estimates 
above are higher than the observed outflow 
speeds in the Seyfert outflows
  ($v_{\infty},_{obs}  \simeq 25 - 150 $ km s$^{-1}$; 
Condon et al. 1996). Nonetheless, according to Eq. 24, 
these observations can be fitted 
by a more compact disk (like that of Fig. 1, for instance) 
with  a somewhat larger ratio
$\dot M_w/\dot M_d$ (Eq. 21). 
 Eqs. 21, 22,  24 and 25 indicate that  disks with 
$r_e \lesssim $  10 pc 
and $\dot M_w(r_i)/\dot M_d \gtrsim $ 0.1 will
 provide $v_{\infty}$ in better agreement with the observed
 Sey outflows. 

In summary,
the results above indicate that the observed terminal
outflow velocities are consistent with an 
outflow generation in the described SB-disk 
scenario with wind mass loss to disk accretion
rate ratios $\gtrsim 0.1$.

One of the attractions of a magnetically driven outflow 
model is that it is 
able not 
only to produce and collimate the outflow, but also to efficiently
extract
 angular 
momentum from the disk, allowing it to collapse (e.g. Blandford 
1976, PP92). Just a small fraction of the infalling mass, 
like that estimated above,
 is required to be carried out by the outflow for it
to take out most of the angular momentum of the system.
Further, in the outflow model investigated here, 
it is easy to show that
the total kinetic power carried by the wind is exactly 
one-half of the gravitational energy released in the
accretion disk, i.e.,
$L_{w} \, = \, {1 \over 2 } \, L_{acc}  \,
 = \, {1 \over 2}  \,   {G  M_{r_e} \dot M_d  \over r_e}  \,
\left[  { 1 - \left( {r_i \over r_e } \right)^{0.45}  } \right] $,
where $M_{r_e}$ is the disk mass within $r_e$. 
Again, using the Arp 220 parametrization, we find that
\begin{equation}
L_w(r_i)  \simeq  1.5 \times 10^{42} {\rm erg s}^{-1}
\left({\dot M_d \over 100 \, M_{\odot} \, {\rm yr}^{-1} }\right)
\left({M_{r_e} \over 2.3 \times 10^9 \, M_{\odot} }\right)
\left({r_e  \over 127 \, {\rm pc} } \right)^{-1}
\left[  { 1 - \left(  {r_i/r_e  \over 0.14 } \right)^{0.45}  } \right] .
\end{equation}
\noindent
This can be compared with the SN energy rate estimated for the 
Arp 220 disk, $L_K \simeq $ 7 $\times$ 10$^{43}$ 
erg s$^{-1}$ (Eq. 4), which results
$L_w/L_K \simeq 2 \times 10^{-2}$, so 
that the total kinetic power carried by the magneto-centrifugally driven 
outflow is much smaller than the  available SN  energy rate 
predicted by the SB models.

The fact that the open field lines must have an inclination 
angle with 
respect  to the disk not larger than $\theta = 70^{o}$ (PP92)
 results 
that 
the height $z_A$ of the Alfv\'en surface is
$z_A \, < \, r_A $ tan $\theta$, or
\begin{equation}
z_A < 150 \, {\rm pc} \,
\left({r_A/r_i \over 3}\right) \,
\left({r_i \over 18 \, pc}\right).
\end{equation}

We can also estimate the outflow number density at the Alfv\'en
surface (and beyond) from the condition that
$ n_A \propto \psi^{2}$ (PP92), and 
according to Eq. 14, 
$ n_A \propto r^{3-\beta \over \alpha} \, \propto \, r^{0.45}$.
Using the number density at the base of the disk, which was 
estimated from the 
Arp 220 disk, $n(r_i) \simeq \, 3 \times 10^{4}$ cm$^{-3}$, the
relation above implies that
\begin{equation}
n_A \, \simeq 5 \times 10^{4} \, {\rm cm^{-3} } \,
\left({r_A/r_i \over 3 }\right)^{0.45} 
\left({r_i \over 18 \, {\rm pc} }\right)^{0.45}.
\end{equation}

According to PP92 models, the flow will recollimate beyond the
fast-magnetosonic  (fm) point. The Mach number at this
point is 
\begin{equation}
M_{fm} = (2^{3/2} \lambda^{1/2} - 3)^{1/2}
\end{equation}

Substituting the value of $\lambda$ given by Eq. (25) we find
  $M_{fm} \, \simeq $ 2.4. Thus flows achieving
modest fm-Mach number $\sim 2$ must recollimate 
and form magnetic focal points beyond about a hight 
$z_A \sim \, 150 $ pc.

\section {Conclusions and Discussion}

SBs with a disk core formed by infalling gas produced during 
star formation in a rotating stellar cluster,  
seem to be able to provide the amount of energy required to
 drive the
weak, directed outflows
of Seyferts galaxies and LIGs. In the latter source class, 
for example,
there seems to be strong observational support for the presence
of nuclear SBs (e.g., SLL98).
We have found that massive nuclear SBs 
with cluster masses
$M_c \, \sim \, 10^{9} \, - \, 10^{10} \, M_{\odot}$, 
disk masses 
$M_d \, \sim \, 10^8 \, - 10^{9} \, M_{\odot}$,
and supernova rates 
$\nu_{SN} \simeq 5 \times 10^{-3} \, - \, 2 $ yr$^{-1}$
(which are consistent with the inferred $\nu_{SN} $
from the observed non-thermal radio power)
may inject kinetic
energies $\sim 10^{41} \, - \, 10^{43} $ erg s$^{-1}$ 
which
are high enough to blow out directed winds from the
accreting disk surface (\S 2),
within the SB lifetimes. 

In our models, the acceleration and 
collimation of the outflow is provided by moderate
magnetic fields 
anchored
into the rotating SB-disk. 
Based on empirical conditions derived from the disk of the prototype
LIG, Arp 220, and the magnetized outflows of 
Sey galaxies and LIGs, 
we have found that 
magnetic fields
with intensities
$\gtrsim 8 \times 10^{-4} \, - 5 \times 10^{-2}$ G 
are able to accelerate 
the flow up to the observed terminal velocities 
($\lesssim$ few 100 km s$^{-1}$ in the case of the Seyferts, and 
$\sim 400-950 $ km s$^{-1}$ in the case of the LIGs), 
with wind mass loss to disk accretion
rate ratios 
$ \dot M_w /\dot M_d \, \gtrsim \, 0.1$ (where
$ \dot M_d \, \sim 100 \, M_{\odot} $ yr$^{-1}$).
The outflow carries only a small fraction (few percent) of the 
available SN kinetic energy rate, 
and is produced within a wind zone in the disk of
 radius  $\lesssim$ 100 pc in the case of the LIGs, and 
$\lesssim$ 10 pc in the case of the Seyfert outflows (\S 3).
The model also predicts a decay of the terminal velocity with radius which is
compatible with observations.

The formation of an outflow in the proposed SB-magnetized disk 
scenario depends on a
series of conditions which are evolutionary  
linked, such as, 
the initial amount of angular momentum and ellipticity in 
the central stellar cluster that will provide the formation of the rotating 
disk of gas from the SB; the development of an 
appropriate magnetic field geometry with open lines 
during disk collaps;
and the ionization state of the gas which must be large enough to
couple the magnetic field to the gas. 
SB-driven outflows are observed not only in IR luminous nuclear 
SB systems, but also in other galaxy types, as for example, blue compacts and 
irregular dwarf galaxies like NGC 1569 (e.g., Greggio et al. 1998).  The  
collimation-acceleration mechanism here described could, in principle, also
apply to the inner regions of  the wide-angled
super-winds observed in these systems. They could also anchor 
a nuclear magnetized disk  generated 
during the SB
if the initial conditions for the SB
were not very much different from those described here.
 However, in the case of the amorphous dwarf systems like 
NGC 1569, which is actually a post-SB system, it is not unlikeley that the 
SB-driven large-scale wind had stripped the dwarf of most of 
its gas before the gas had 
time to partially collect into the center and develop a nuclear rotating disk and
a more collimated, coherent outflow portion.
In forthcoming work, we aim
to explore
self-consistent numerical models to examine the dynamical evolution
 of outflows produced by a rotating magnetized disk originated 
in different  SB environments and determine their range of applicability. 
For that purpose, we will
employ a modified version of a 3-D numerical SPH code 
originally designed for  supersonic jet studies (e.g., 
de Gouveia Dal Pino \& Benz 1993, 
de Gouveia Dal Pino \& Birkinshaw 1996, 
Cerqueira \& de Gouveia Dal Pino 1998).

Pure hydrodynamic models (e.g., Suchkov et al. 1994) make specific 
predictions for the observed abundance pattern in the SB-driven 
super-winds. 
In those models, the material at the wall of the 
vertical cavity, for example, is mostly composed of 
unprocessed ISM gas
that has been shock-heated by the emerging outflow, while the
enriched material that is ejected by the SN  should 
fill the interior of the cavity. Similarly, as a collimated
 magnetized outflow
impinges supersonically the ambient medium, it 
must create a bow shock pattern at the outer edge (see, e.g.,
Cerqueira, de Gouveia Dal Pino, \& Herant 1997)
where most of the shocked material that produces the emission lines
is made of ISM gas with no significant abundance 
enrichment. The observed optical emission lines should then trace 
normal abundances just like predicted in hydrodynamical models. The hotter,
enriched material that is ejected by the SN
should emerge mainly along with  the collimated outflow. 

We note that in our disk model the magnetic pressure
in the disk may  exceed the gas pressure. This may cause the disk to
become unstable to magnetic buoyancy. In such a case, the magnetic
field lines may arise in loops from the disk surface and undergo 
reconnection. This could in turn provide the formation of hot 
filamentary
and clumpy structure on the disk corona. It is interesting to
note that a $patchy$ disk corona is required by the 
X-ray and $\gamma$-ray observations of the nuclear region 
of Sey galaxies in order to explain their complex high-energy
spectra (e.g., Johnson et al. 1997).

Finally, we should inquire about the fate of the SB disk (see also GT97). 
We have seen that it may rapidly become self-gravitating
(\S 2.2).
As a consequence,  it is expected to suffer fragmentation and 
star formation. Either, the disk may become bar unstable, 
a condition which is met if $\delta_b \, = \, T_{rot}/W > 0.14$, 
where 
$T_{rot}$ is the rotational energy and 
$W$ is the gravitational potential energy of the disk 
(e.g. Shlosman et al. 1989).
Our SB disks result $\delta_b \, = \, 0.5$, so that they can, in 
principle, 
become  unstable to bar formation and undergo fission 
into two components 
which eventually may collapse to a supermassive 
BH (Begelman, Blandford, \& Rees 1980).
Based on the fact that the formation of BHs seems to be plausible 
in a SB, we can speculate that both, 
the standard BH and the SB mechanisms may be occurring in 
different stages of the AGN evolution.
In this evolutionary scenario, the stronger more collimated jets 
of radio loud QSOs, for example, could be produced in the 
higher-level 
activity phase of the BH accreting the remains of the
disk material. This hypothesis  is 
examined elsewhere.

\acknowledgements
This work has been partially supported by grants of the Brazilian
agencies FAPESP and CNPq. 
E.M.G.D.P. acknowledges the kind hospitality of the staff of 
the Abdus Salam International Centre for Theoretical Physics,
in Trieste, where this work was accomplished. We are also 
indebted to the referee for his comments and suggestions that 
we believe have helped to improve this work.

\newpage

{\bf Figure Caption}

Figure 1. a) Evolution of a stellar cluster  and its core 
disk, with a Salpeter IMF  
$(M_l \,= \, 1 M_{\odot}$, 
$M_u \, = \, 120 M_{\odot}$), $N_c \, = \, 3 \times 10^8$,
$r_c \, = \, 53$ pc, 
 total cluster mass $M_c \, = \, 9.5 \times 10^8 
M_{\odot}$,
$R_d \, = \, 2.6$ pc, 
 total number of stars in the disk 
$N_d \, = \, 9 \times 10^7$,
and total stellar mass in the disk
$M_d \, = \, 1.6 \times 10^8 
M_{\odot}$.
Luminosity sources shown: dotted - supernovae in the cluster and 
disk;
solid - rate of energy released by the infalling gas in the disk;
dashed - starlight in the cluster and disk; and
dot-dashed - the Eddington luminosity limit in the disk. 
b) The corresponding rates of supernovae production 
$\nu_{SN}$ 
in the cluster
and the disk.
For a system with masses 
and radii about ten times 
larger (smaller) than those above,
the luminosity values and supernova rates increase (decrease) by 
an order of magnitude.

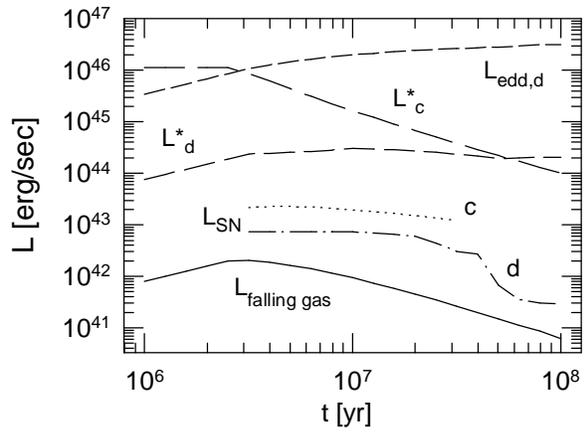
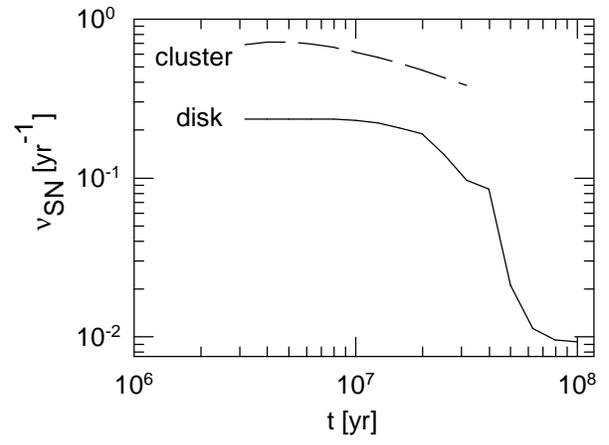
\end{document}